\documentclass [useAMS]{mn2e}
\usepackage {graphicx}

\begin{document}

\title{Does dichotomy of active galactic nuclei only depend on black hole
spins?}
\author[]{Yong-Chun Ye, Ding-Xiong Wang$^{*}$ \\
$$ Department of Physics, Huazhong University of Science and Technology, Wuhan,430074,China \\
$^*$Send offprint requests to: D.-X. Wang (dxwang@hust.edu.cn) }
\maketitle

\begin{abstract}

A toy model for jet powers and radio loudness of active galactic
nuclei (AGNs) is proposed based on the coexistence of the
Blandford-Znajek (BZ) and magnetic coupling (MC) processes
(CEBZMC) in black hole (BH) accretion disc. It turns out that both
the jet powers and radio loudness of AGNs are controlled by more
than one physical parameter besides the BH spin. The observed
dichotomy between radio-loud and radio-quiet AGNs is well
interpreted by the two parameters, the BH spin and the power-law
index of the variation of the magnetic field on the disc.
Furthermore, we discuss the correlation of jet powers with radio
loudness of AGNs in terms of the two parameters. It is found that
the contours of radio loudness are approximately in accord with
those of jet powers for several 3CR radio sources, implying
roughly that the stronger jet powers corresponds to stronger radio
loudness. In addition we discuss the correlation of the jet powers
and radio loudness of AGNs with the position of the inner edge of
an accretion disc. These results imply that the ``spin paradigm''
for radio loudness of AGNs might be modified by a scenario
containing more physical parameters.

\end{abstract}



\section{INTRODUCTION}

As is well known, active galactic nuclei (AGNs) are characterized
not only by the large range and bimodal distribution of their
radio loudness, but also by jets of powering large-scale radio
structure. It has been observed that AGNs can approximately be
divided into two classes: radio-loud (RL) AGNs and radio-quiet
(RQ) AGNs. In 1989, Kellerman et al. indicated that RL AGNs are
rather rare (they consist only about 10{\%} of the AGN population)
and they almost reside in elliptical galaxies, having the ratio of
their radio to optical fluxes ${F_{5GHz} } \mathord{\left/
{\vphantom {{F_{5GHz} } {F_B }}} \right.
\kern-\nulldelimiterspace} {F_B } > 10$. Recently, Cirasuolo et
al. (2003) suggested that RL AGNs consist about 5{\%} of the AGN
population or more less.

Blandford (1990) proposed the ``spin paradigm'' that the key
parameter which determines the jet powers and radio loudness of
AGNs is the black hole spin. Wilson {\&} Colbert (1995) argued
that high spin BHs exist only if formed by the coalescence of two
BHs, which would take place mostly in the environment of
elliptical galaxies. Considering the fact that a BH is spun down
much more effectively in retro-grade disc accretion than the
pro-grade case, Moderski et al. (1997, hereafter M97) proposed a
model of switching randomly between pro- and retro-grade disc
accretion with the Blandford-Znajek (BZ, 1977) process to explain
the observed radio dichotomy of AGNs. Phenomenologically, RL AGNs
are always associated with large-scale radio jets and lobes, while
the RQ sources have very little or weak radio-emitting ejecta.
This phenomenon means that there may exist a dramatic relation
between jet powers and radio loudness of AGNs.

The strength of the BZ mechanism is related closely to accretion
mode. Recently, Cao and Rawlings (2004) argued that perhaps most
of the 3CR FR I radio galaxies have normal accretion discs rather
than advection dominated accretion flows or adiabatic
inflow-outflow solution flows, otherwise the BZ mechanism cannot
provide sufficient power to explain their high radio luminosities.

On the other hand, the strength of the BZ mechanism depends on the
configuration of the magnetic field in BH accretion disc.
Recently, much attention has been paid on the magnetic coupling
(MC) of a BH with its surrounding disc, and the recent
\textit{XMM-Newton} observation of a broad Fe $K\alpha $ line in
the bright Seyfert 1 galaxy MCG-6-30-15 suggested strongly the
existence of the MC effects (Wilms et al. 2001; Li 2002; Wang et
al. 2003a, 2003b, hereafter W03a and W03b, respectively). In fact
the MC process can be regarded as a variant of the BZ process, in
which energy and angular momentum are transferred from a
fast-rotating BH to its surrounding disc. In addition, it is shown
that the BZ and MC processes might coexist, provided that the BH
spin and the power-law index are greater than some critical values
(W03b).

In this paper we intend to clarify the correlation between jet
powers and radio loudness of AGNs in the model of the coexistence
of the BZ and MC processes (CEBZMC) on the basis of the normal
accretion discs, assuming that the jet is powered in the BZ
process, and the energy transferred from a BH to the disc in the
MC process contributes the bolometer luminosity of the disc as
well as disc accretion. It turns out that both jet powers and
radio loudness of AGNs do not depend on the BH spin only, but on
the other physical parameters probably.

This paper is organized as follows. In Section 2 radio loudness of
AGNs is discussed in a parameter space consisting of the BH spin
and the power-law index, and a reasonable explanation for the
observed radio dichotomy of AGNs is given. In Section 3, the
contours of jet powers of AGNs are analyzed, and the correlation
of jet powers with radio loudness of AGNs is discussed also in the
parameter space. Finally, in Section 4, we discuss the correlation
of the jet powers and radio loudness of AGNs with the position of
the inner edge of an accretion disc, and summarize our main
results. Throughout this paper the geometric units $G = c = 1$ are
used.


\section{RADIO DICHOTOMY OF AGNS BASED ON CEBZMC}

Since the configuration of magnetic field in BH magnetosphere is
very complicated, we discussed magnetic extraction of energy from
a rotating BH based on several simplified assumptions (Wang et al.
2002, hereafter W02). Following Blandford (1976) the poloidal
magnetic field is assumed to vary as a power law with the radial
coordinate $\xi $ of the disc,


\begin{equation}
\label{eq1} B_D^p = B_H^p \left[ {{r_H } \mathord{\left/
{\vphantom {{r_H } {\varpi _D \left( {r_{ms} } \right)}}} \right.
\kern-\nulldelimiterspace} {\varpi _D \left( {r_{ms} } \right)}}
\right]\xi ^{ - n} \quad , \quad 1 < \xi < \xi _{out} ,
\end{equation}

\noindent where $B_D^p $ and $B_H^p $ is the poloidal component of
the magnetic field on the disc and on the horizon respectively,
and $n$ is the power-law index of $B_D^p $ varying with $\xi $,
and $\xi \equiv r \mathord{\left/ {\vphantom {r {r_{ms} }}}
\right. \kern-\nulldelimiterspace} {r_{ms} }$ is defined as a
radial parameter in terms of $r_{ms} $. The quantities $r_H $ and
$\varpi _D \left( {r_{ms} } \right)$ are the radius of the horizon
and the cylindrical radius at the inner edge of the disc,
respectively.

In W02 we proposed a configuration of magnetic field as shown in
Fig.1, and the expressions for the BZ and MC power in CEBZMC are
given as follows (W02; W03a; W03b).


\begin{equation}
\label{eq2} {P_{BZ} } \mathord{\left/ {\vphantom {{P_{BZ} } {P_0
}}} \right. \kern-\nulldelimiterspace} {P_0 } = 2a_ * ^2
\int_0^{\theta _M } {\frac{k\left( {1 - k} \right)\sin ^3\theta
d\theta }{2 - \left( {1 - q} \right)\sin ^2\theta }} , \quad 0 <
\theta < \theta _M ,
\end{equation}


\begin{equation}
\label{eq3} {P_{MC} } \mathord{\left/ {\vphantom {{P_{MC} } {P_0
}}} \right. \kern-\nulldelimiterspace} {P_0 } = 2a_ * ^2
\int_{\theta _M }^{\theta _L } {\frac{\beta \left( {1 - \beta }
\right)\sin ^3\theta d\theta }{2 - \left( {1 - q} \right)\sin
^2\theta }} , \quad \theta _M < \theta < \theta _L ,
\end{equation}

\noindent where the parameter $q \equiv \sqrt {1 - a_ * ^2 } $ is
a function of the BH spin, $k \equiv {\Omega _F } \mathord{\left/
{\vphantom {{\Omega _F } {\Omega _H }}} \right.
\kern-\nulldelimiterspace} {\Omega _H }$ is the ratio of the
angular velocity of the open field lines to that of the horizon.
Usually, $k = 0.5$ is taken for the optimal BZ power (Macdonald
{\&} Thorne 1982). And $\beta $ is the ratio of the angular
velocity of the closed field lines to that of the horizon, which
depends on the BH spin and the place where the field line
penetrates on the disc. The parameter $\beta $ is related to $a_ *
$ and $\xi $ by


\begin{equation}
\label{eq4} \beta \equiv {\Omega _D } \mathord{\left/ {\vphantom
{{\Omega _D } {\Omega _H }}} \right. \kern-\nulldelimiterspace}
{\Omega _H } = \frac{2\left( {1 + q} \right)}{a_ * }\left[ {\left(
{\sqrt \xi \chi _{ms} } \right)^3 + a_ * } \right]^{ - 1} \quad .
\end{equation}

\noindent In equation (\ref{eq3}) $P_0 $ is defined by


\begin{equation}
\label{eq5} P_0 \equiv \left( {B_H^p } \right)^2M^2 \approx B_4^2
M_9^2 \times 6.59\times 10^{46}erg \cdot s^{ - 1},
\end{equation}

\noindent where $B_4 $ is the poloidal magnetic field at the
horizon in the unit of $10^4Gauss$, and $M_9 $ is the BH mass in
the unit of $10^9M_ \odot $. The half-opening angle $\theta _M $
is the angular boundary between the open and closed field lines on
the BH horizon, being determined by the following mapping
relation,


\begin{equation}
\label{eq6} \cos \theta _M - \cos \theta _L = \int_1^{\xi _{out} }
{\mbox{G}\left( {a_ * ;\xi ,n} \right)d\xi } ,
\end{equation}

\noindent where


\begin{equation}
\label{eq7} \begin{array}{l} G\left( {a_ * ;\xi ,n} \right) = \\
\\
\quad\quad \frac{\xi ^{1 - n}\chi _{ms}^2 \sqrt {1 + a_ * ^2 \chi
_{ms}^{ - 4} \xi ^{ - 2} + 2a_ * ^2 \chi _{ms}^{ - 6} \xi ^{ - 3}}
}{2\sqrt {\left( {1 + a_
* ^2 \chi _{ms}^{ - 4} + 2a_ * ^2 \chi _{ms}^{ - 6} }
\right)\left( {1 - 2\chi _{ms}^{ - 2} \xi ^{ - 1} + a_ * ^2 \chi
_{ms}^{ - 4} \xi ^{ - 2}} \right)} }.
\end{array}
\end{equation}

One of the important results in the model of CEBZMC is that the
angle $\theta _M $of the magnetic flux tube on the horizon is
determined by $a_ * $ and $n$.

\begin{figure}
\vspace{0.5cm}
\begin{center}
\includegraphics[width=6cm]{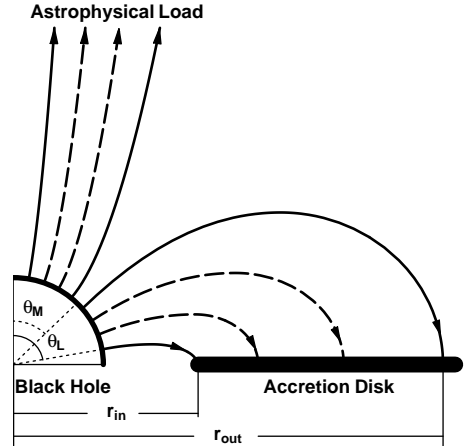}
\caption{The poloidal magnetic field connecting a rotating BH with
remote load and a disc} \label{fig1}
\end{center}
\end{figure}

Radio loudness of AGNs is defined as the radio-to-optical flux
ratio $F_{5GHz} / F_B $ (Kellerman et al. 1989). Assuming that the
fraction of jet power converted to radiation is 10{\%}, and that
bolometric correction for jet radiation at $v_R \sim 5GHz$ is of
the same order as that for the disc radiation in the B-band, M97
expressed the radio loudness of AGNs as the ratio $P / L$, where
$P$ is the total radio luminosity powered by the BZ process, and
$L$ is the total bolometer luminosity of the accretion disc. In
our model we suggest that the total radio luminosity is still
powered by the BZ process, while the total bolometer luminosity of
the disc is contributed by the MC process as well as the disc
accretion, i.e.,


\begin{equation}
\label{eq8} P = P_{BZ} , \quad L = (1 - E_{ms} )\dot {M}_D +
P_{MC} ,
\end{equation}

\noindent where $\dot {M}_D $ is accretion rate, and $E_{ms} $ is
specific energy corresponding to the radius of the marginally
stable orbit $r_{ms} $ and it reads (Novikov {\&} Thorne 1973)


\begin{equation}
\label{eq9} E_{ms} = {\left( {1 - 2\chi _{ms}^{ - 2} + a_ * \chi
_{ms}^{ - 3} } \right)} \mathord{\left/ {\vphantom {{\left( {1 -
2\chi _{ms}^{ - 2} + a_ * \chi _{ms}^{ - 3} } \right)} {\left( {1
- 3\chi _{ms}^{ - 2} + 2a_ * \chi _{ms}^{ - 3} } \right)^{1 /
2}}}} \right. \kern-\nulldelimiterspace} {\left( {1 - 3\chi
_{ms}^{ - 2} + 2a_ * \chi _{ms}^{ - 3} } \right)^{1 / 2}}.
\end{equation}

As the magnetic field on the BH is supported by the surrounding
disc, there are some relations between $B_H $ and $\dot {M}_D $.
In this paper, we take the relation suggested in M97, arising from
the balance between the pressure of the magnetic field on the
horizon and the ram pressure of the innermost parts of an
accretion flow, i.e.,


\begin{equation}
\label{eq10} \begin{array}{l}{2\dot {M}_D } \mathord{\left/
{\vphantom {{2\dot {M}_D } {\left( {1 + q} \right)^2}}} \right.
\kern-\nulldelimiterspace} {\left( {1 + q} \right)^2} = \left(
{B_H^p } \right)^2M^2 \equiv P_0.
\end{array}
\end{equation}

\noindent Incorporating equations (\ref{eq8}) and (\ref{eq10}), we
express radio loudness as


\begin{equation}
\label{eq11} \frac{P}{L} = \frac{2P_{BZ} / P_0 }{(1 - E_{ms} )(1 +
q)^2 + 2P_{MC} / P_0 }.
\end{equation}

\noindent Incorporating equations (\ref{eq2}), (\ref{eq3}),
(\ref{eq9}) and (\ref{eq11}), we have the curves of radio loudness
of AGNs versus $a_ * $ for the given $n$ and those versus $n$ for
the given $a_ * $ as shown in Figs.2a and 2b, respectively.

\begin{figure}
\vspace{1.4cm}
\begin{center}
{\includegraphics[width=6cm]{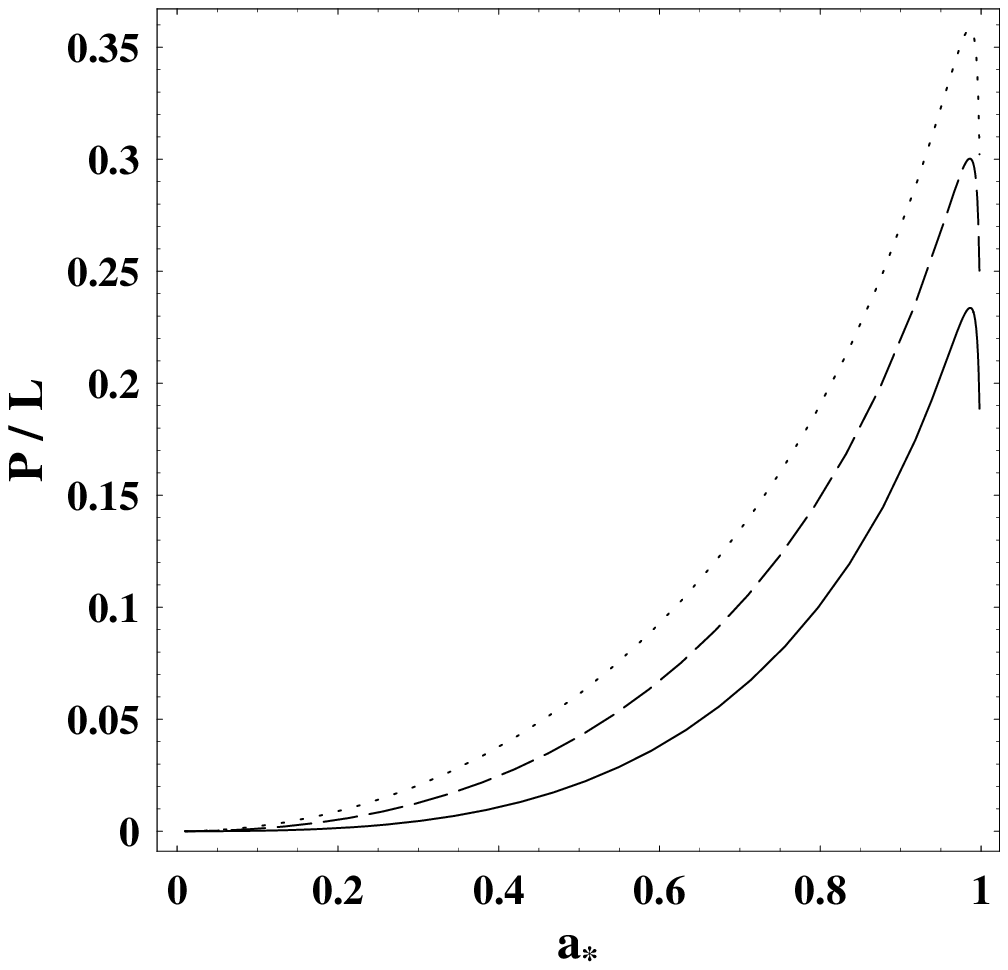}
 \centerline{\quad\quad\quad(a)}
 \includegraphics[width=6cm]{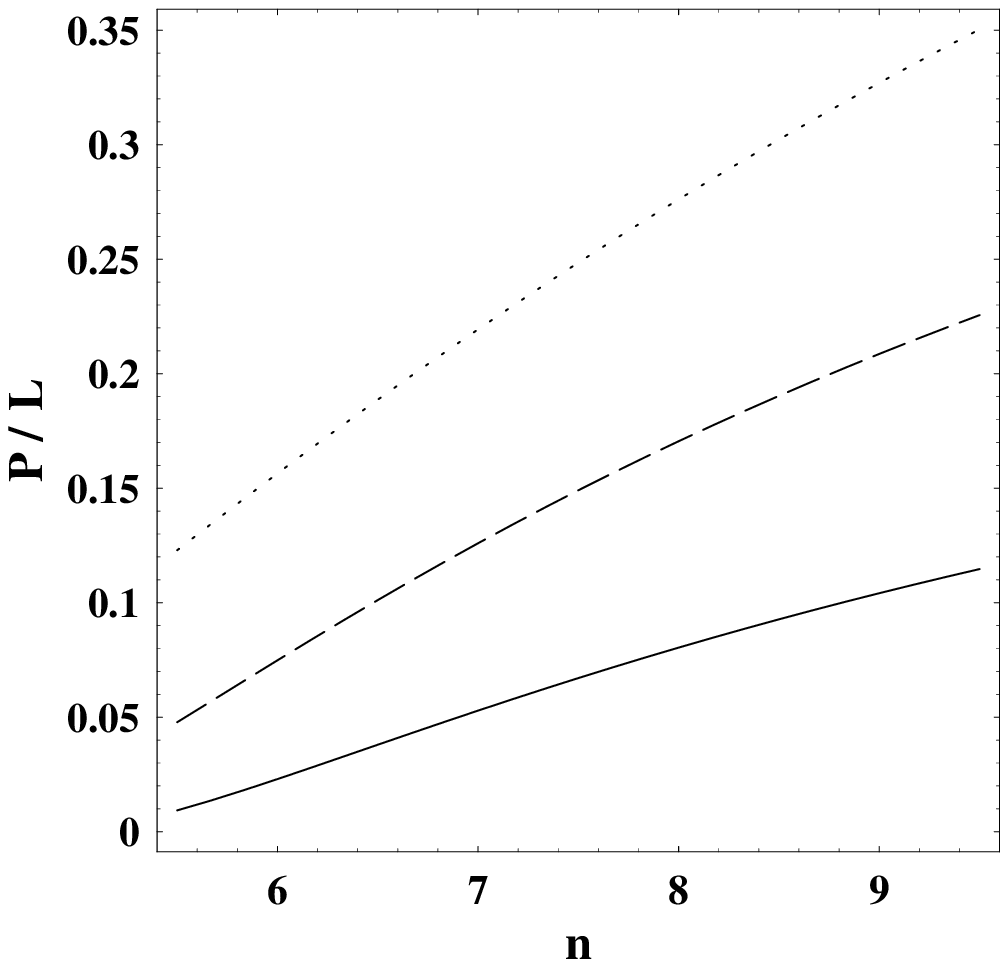}
 \centerline{\quad\quad\quad(b)}}
 \caption{Curves of radio loudness $P / L$ (a)
versus $a_\ast $ with $n$=6.5, 7.5 and 8.5, and (b) versus $n$
with $a_\ast $=0.6, 0.8 and 0.998 in solid, dashed and dotted
lines, respectively.} \label{fig2}
\end{center}
\end{figure}

Radio loudness of AGNs varies non-monotonically with the
increasing $a_ * $ and the constant $n$, attaining its maximum as
the BH spin approaches unity as shown in Fig.2a, while it
increases monotonically with the increasing $n$ and the constant
$a_ * $ as shown in Fig.2b.

By using equation (\ref{eq11}) we have the contours of radio
loudness of AGNs in the parameter space consisting of $a_
* $ and $n$ as shown in Fig.3, where the value of $P / L$ is
labeled beside each contour, and the thick solid line labeled
0.001 is regarded as the dividing line between RQ and RL AGNs
(M97).

\begin{figure}
\vspace{0.5cm}
\begin{center}
\includegraphics[width=6cm]{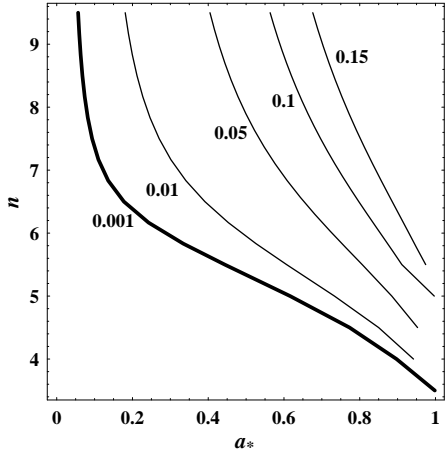}
\caption{ The contours corresponding to radio loudness with $P /
L$= 0.001, 0.01, 0.05, 0.1 and 0.15 in $a_ * - n$ parameter
space.} \label{fig3}
\end{center}
\end{figure}

Following M97, in the ``spin paradigm'' the total radio luminosity
is powered by the BZ process, and the total bolometer luminosity
of the disc only arises from disc accretion without the MC
process, and the radio loudness of AGNs is given by


\begin{equation}
\label{eq12} (P / L)_{M97} = \frac{a_\ast ^2 }{64(1 - E_{ms} )},
\end{equation}

\noindent where we take


\begin{equation}
\label{eq13} P = P_{BZ} = \frac{k\left( {1 - k} \right)}{16}a_ *
^2 \dot {M}_D  \quad\quad with \quad\quad k = 0.5,
\end{equation}

\noindent and


\begin{equation}
\label{eq14} L = (1 - E_{ms} )\dot {M}_D .
\end{equation}

\noindent Thus $(P / L)_{M97} $ only depends on the BH spin, and
the curve of its variation with $a_ * $ is shown in Fig.4.

\begin{figure}
\vspace{0.5cm}
\begin{center}
\includegraphics[width=6cm]{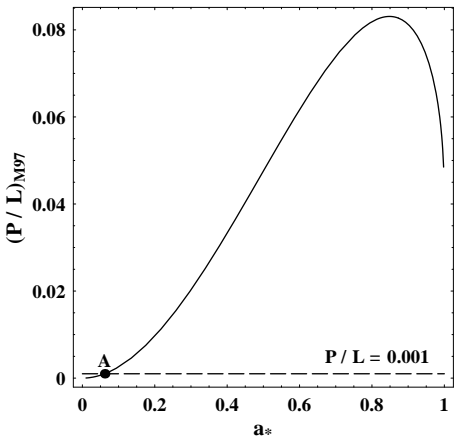}
\caption{ Radio loudness $(P / L)_{M97} $ versus $a_\ast $.}
\label{fig4}
\end{center}
\end{figure}

From Fig.4 we find that $(P / L)_{M97} $ also varies
non-monotonically with the increasing $a_ * $, and the curve of
radio loudness of AGNs (solid line) intersects with the critical
curve of radio loudness $P / L = 0.001$ (dashed line) at point
\textbf{A} (0.06, 0.001), implying a very low BH spin ($a_ * <
0.06)$ is required for RQ AGNs in M97. Considering the fact that
BHs could be easily spun up in disc accretion at later epochs
(Bardeen 1970), the conclusion that low-spin dominated states of
BHs could be questionable.

In contrast to M97, radio loudness of AGNs in our model depends on
both the BH spin $a_ * $ and the power-law index $n$: RL AGNs
require high values of both $a_ * $ and $ n$, and RQ AGNs might
correspond to either high or low BH spin, provided that the value
of the power-law index $ n$ is low enough. Dependence of radio
loudness of AGNs on the values of $a_ * $ and $ n$ is listed in
Table 1. It seems reasonable that the probability of AGNs in CASE
A should be much less than the sum of the probabilities in CASE B,
C and D, and this might provide a possible explanation for the
observations that the RL AGNs consist very small fraction of the
AGN population.

\begin{table*}
\caption{Dependence of radio loudness of AGNs on the parameters
$a_\ast $ and $n$.}
\begin{tabular}
{|p{90pt}|p{90pt}|p{90pt}|p{90pt}|} \hline CASE& BH spin $a_\ast
$& Power-law index $n$&
Radio loudness \\
\hline A& High& High&
RL  \\
\hline B& High& Low&
RQ \\
\hline C& Low& High&
RQ \\
\hline D& Low& Low&
RQ \\
\hline
\end{tabular}
 \label{tab1}
\end{table*}

\section{CORRELATION OF JET POWERS WITH RADIO LOUDNESS OF
AGNS BASED ON CEBZMC}

If the BZ mechanism is regarded as the major mechanism for
powering radio jets in AGNs (Blandford {\&} Znajek 1977; Rees
1984), the dimensionless jet power can be written as


\begin{equation}
\label{eq15} \tilde {P}_{jet} \left( {a_ * ,n} \right) = P_{jet} /
P_0 = \frac{1}{2}a_ * ^2 \int_0^{\theta _M } {\frac{\sin ^3\theta
d\theta }{2 - \left( {1 - q} \right)\sin ^2\theta }} ,
\end{equation}

\noindent where equation (\ref{eq15}) is derived by taking $k =
0.5$ in equation (\ref{eq2}). By using equation (\ref{eq15}), we
have the curves of jet power $\tilde {P}_{jet} $ versus $a_\ast $
for the given $n$ and those versus $n$ for the given $a_\ast $ as
shown in Figs.5a and 5b, respectively.

\begin{figure}
\vspace{1.4cm}
\begin{center}
{\includegraphics[width=6cm]{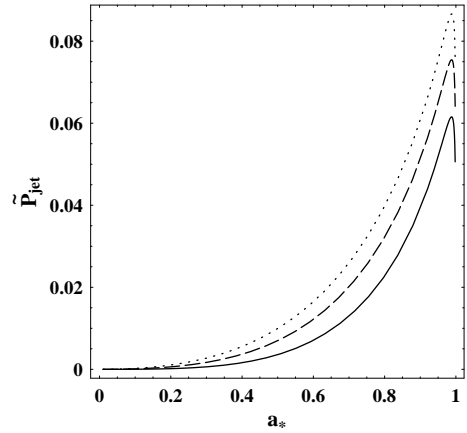}
 \centerline{\quad\quad\quad(a)}
 \includegraphics[width=6cm]{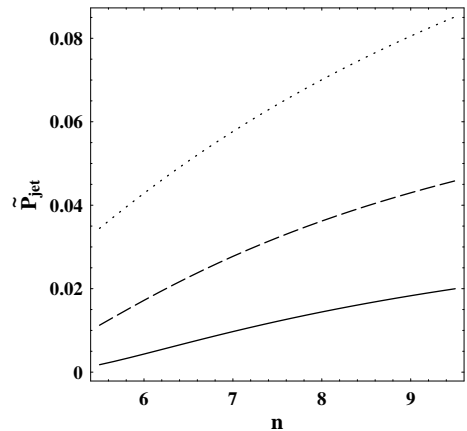}
 \centerline{\quad\quad\quad(b)}}
 \caption{Jet power $\tilde {P}_{jet} $ (a)
versus $a_\ast $ with $n$=6.5, 7.5 and 8.5 in solid, dashed and
dotted lines, respectively; (b) versus $n$ with $a_\ast $=0.6, 0.8
and 0.998 in solid, dashed and dotted lines, respectively. }
\label{fig5}
\end{center}
\end{figure}

Comparing Fig.5 with Fig.2, we find that the curves of $\tilde
{P}_{jet} $ are very analogous to those of radio loudness of AGNs,
and this similarity motivates us to study the correlation these
two quantities based on the model of CEBZMC. The values of the BH
mass, the poloidal magnetic field on the horizon, absolute jet
powers and radio loudness of some 3CR radio sources are listed in
Table 2, where the poloidal magnetic fields on the horizon, $B_H^P
$, are estimated by the following relation (Beskin 1997),


\begin{equation}
\label{eq16} B_H^P \sim 10^4M_9^{{ - 1} \mathord{\left/ {\vphantom
{{ - 1} 2}} \right. \kern-\nulldelimiterspace} 2} Gauss
\end{equation}

\begin{table*}
\caption{BH mass, poloidal magnetic field on the horizon, jet
powers and radio loudness of some 3CR radio sources: 3C345, 3C273,
3C390.3 and 3C29.}
\begin{tabular}
{|p{40pt}|p{40pt}|p{35pt}|p{46pt}|p{56pt}|p{50pt}|p{35pt}|p{40pt}|p{35pt}|}
\hline Objects \par (\ref{eq1})& Mass \par (2a)& Ref \par (2b)&
$B_H^P $ \par (\ref{eq3})& $P_{jet} $ \par (4a)& $\tilde {P}_{jet}
$
\par (4b)& Ref \par (4c)& Remark \par (5a)&
Ref \par (5b) \\
\hline 3C345& 0.82& WHS& 1.1& $3.72\times 10^{45}$& $6.86\times
10^{ - 2}$& CPG& RL&
WHS \\
\hline 3C273& 1.82& WHS& 0.74& $6.61\times 10^{45}$& $5.51\times
10^{ - 2}$& CPG& RL&
WHS \\
\hline 3C390.3& 0.26& MCF& 1.97& $1.05\times 10^{44}$& $6.18\times
10^{ - 3}$& CPG& RL&
WWNR \\
\hline 3C29& 1.26& CR& 0.89& $3.16\times 10^{44}$& $3.81\times
10^{ - 3}$& CR& RL&
XLB \\
\hline
\end{tabular}
 \label{tab2}
\end{table*}
\begin{table*}
\begin{minipage}{170mm}
Notes for the table 2. Column (\ref{eq1}): source name, Column
(2a): the BH mass ($\times 10^9M_ \odot )$, Column (2b): reference
for the BH mass, Column (\ref{eq3}): the poloidal magnetic field
on the horizon ($\times 10^4Gauss)$, Column (4a): jet power of the
source ($erg \cdot s^{ - 1})$, Column (4b): dimensionless jet
power of the source, Column (4c): reference for the jet power,
Column (5a):
remarks:RL=radio-loud, Column (5b): reference for the remarks. \\
References: (CR) Cao et al. 2004; (XLB) Xu et al. 1999; (WHS) Wang
et al. 2003; (CPG) Celotti et al. 1997; (MCF) Marchesini et al.
2004; (WWNR) Willem et al. 1997.
 \end{minipage}
\end{table*}

Combining equations (\ref{eq11}) and (\ref{eq15}) with the data in
Table 2, we have the contours of $\tilde {P}_{jet} $
(dimensionless jet power) and the dividing line ($P / L = 0.001$
in thick solid line) of radio loudness between RQ and RL AGNs in
$a_ * - n$ parameter space as shown in Fig.6.

\begin{figure}
\vspace{0.5cm}
\begin{center}
\includegraphics[width=6cm]{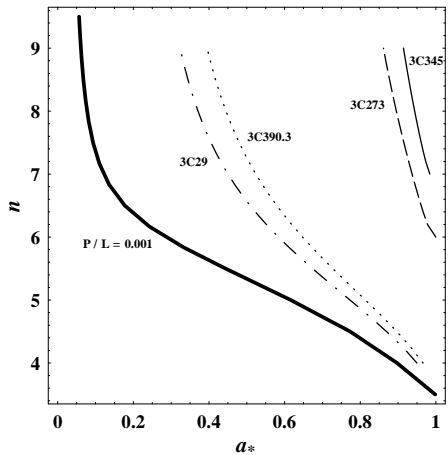}
\caption{The dividing line of radio loudness with $P / L = 0.001$,
and the contours of $\tilde {P}_{jet} $ of 3C345, 3C273, 3C390.3
and 3C29 in solid, dashed, dotted and dash-dotted lines,
respectively.} \label{fig6}
\end{center}
\end{figure}

Inspecting Table 2 and Fig.6, we find the following correlation of
$\tilde {P}_{jet} $ with radio loudness of AGNs. The contour of
$\tilde {P}_{jet} $ with larger value is farther away from the
dividing line of radio loudness in the top right part of the
parameter space than that with smaller value, and this result
implies that the greater $\tilde {P}_{jet} $ generally corresponds
to the greater radio loudness.

This correlation can be clarified more concretely in $a_ * - n$
parameter space as shown in Fig.7. We find that the contour of
radio loudness $P / L = 0.025$ (solid line) intersects with the
dash-dotted line at point A (0.42, 7.33) and the dotted line at
point B (0.83, 4.92), respectively. These two intersections imply
the following correlation of $\tilde {P}_{jet} $ with radio
loudness: A quasar with stronger $\tilde {P}_{jet} $ may has
larger radio loudness than a quasar with weaker $\tilde {P}_{jet}
$ for a very wide value ranges of the BH spin $a_ * $ and the
power-law index $n$, and the counter example occurs only for very
extreme values of the two parameters. For example, quasar 3C 390.3
should have larger radio loudness than quasar 3C 29 for most
values of $a_ * $ and $n$, and the order could be inversed only if
both of the following value ranges of $a_ * $ and $n$ are
satisfied: (i) $a_
* < 0.42$ with $n > 7.33$ for 3C 29, and (ii) $a_ * > 0.83$ with $n < 4.92$
for 3C 390.3. This argument for the correlation could be
applicable to other AGNs, and the correlation of $\tilde {P}_{jet}
$ with radio loudness of AGNs might be a conjecture to be tested
in future.

\begin{figure}
\vspace{0.5cm}
\begin{center}
\includegraphics[width=6cm]{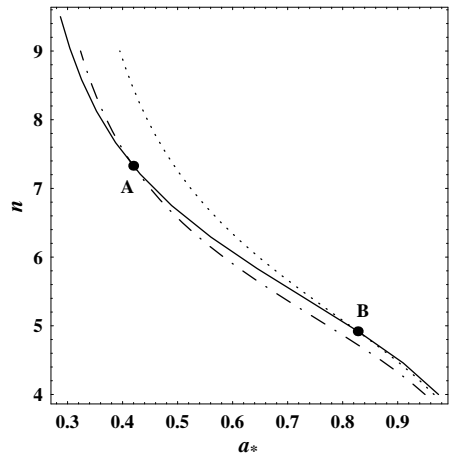}
\caption{The contour of radio loudness with $P / L = 0.025$, and
the contours of $\tilde {P}_{jet} $ of 3C390.3 and 3C29 in dotted
and dash-dotted lines, respectively.} \label{fig7}
\end{center}
\end{figure}

\section{DISCUSSION}

In this paper we discuss the correlation of the jet powers and
radio loudness of AGNs based on the model of CEBZMC. It turns out
that both radio loudness and jet powers of AGNs probably depend on
more than one physical parameter. An explanation for the observed
radio dichotomy of AGNs is given, which is difficult to be
understood in the ``spin paradigm''.

\begin{figure}
\vspace{0.5cm}
\begin{center}
{\includegraphics[width=6cm]{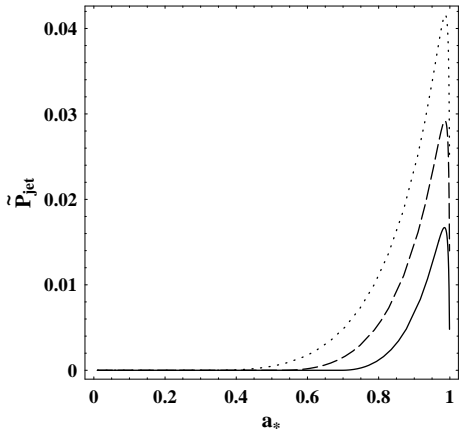}
 \centerline{\quad\quad\quad(a)}
 \includegraphics[width=6cm]{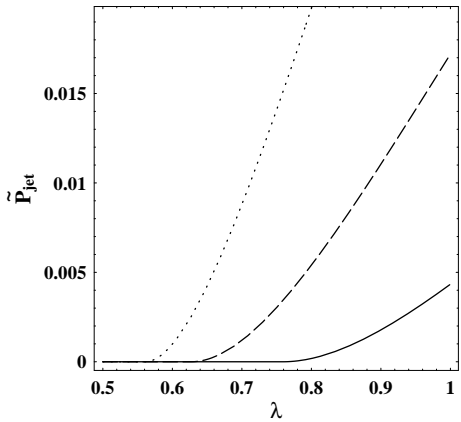}
 \centerline{\quad\quad\quad(b)}}
 \caption{Jet power $\tilde {P}_{jet} $ in the case of a
thick disc (a) versus $a_\ast $ with $\lambda $=0.7, 0.8 and 0.9
in solid, dashed and dotted lines, respectively; (b) versus
$\lambda $ with $a_\ast $=0.6, 0.8 and 0.998 in solid, dashed and
dotted lines, respectively. } \label{fig5}
\end{center}
\end{figure}

\begin{figure}
\vspace{0.5cm}
\begin{center}
\includegraphics[width=6cm]{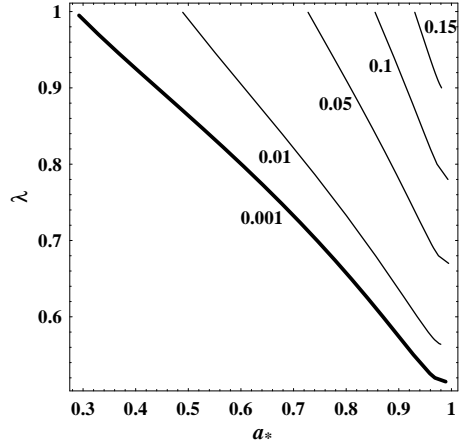}
\caption{ The contours corresponding to radio loudness in the case
of a thick disc with $P / L$= 0.001, 0.01, 0.05, 0.1 and 0.15 in
$a_
* - \lambda $ parameter space.} \label{fig3}
\end{center}
\end{figure}

As is well known, the jet power driven by the BZ mechanism is
proportional to square of the BH spin, and the total bolometer
luminosity of the disc is also related closely to the BH spin by
equation (\ref{eq14}). These relations are essential to the ``spin
paradigm'', and a scenario of switching randomly between pro- and
retro-grade disc accretion is suggested to depress the BH spin in
M97. Unfortunately, we have not found reliable evidence about this
random switching of accretion modes in AGNs.

Compared with M97, our model for radio loudness and jet powers of
AGNs is based on CEBZMC, in which the MC of a BH with its
surrounding disc is taken into account. It is the MC mechanism
that involves the closed field lines connecting the rotating BH
with the disc, and the power-law index $n$ is related to the
variation of the magnetic field expressed by equation (\ref{eq1}).
The index $n $ indicates the degree of the magnetic field is
concentrated in the inner disc, which depends probably on the
distribution of the current on the disc. The calculations in our
model show that the index $n$ plays very important role in
determining radio loudness and jet powers of AGNs as well as the
BH spin.

In fact, both jet powers and radio loudness of AGNs could be
related to other parameters. As is well known, the inner edge of a
thick disc is located between $r_{mb} $ and $r_{ms} $ (Abramowicz
et al. 1978, Abramowicz {\&} Lasota 1980),

\begin{equation}
\label{eq17} r_{mb} \le r_{in} < r_{ms} ,
\end{equation}

\noindent where $r_{mb} $ and $r_{in} $ are the radii of the
innermost bound circular orbit and inner edge of a thick disc,
respectively.

\begin{equation}
\label{eq18}  r_{mb} = M\left( {1 + \sqrt {1 - a_ * } } \right)^2.
\end{equation}

\noindent The radius $r_{in} $ can be expressed in terms of
$r_{mb} $ and $r_{ms} $ by introducing a parameter $\lambda $ as
follows (Wang 1999).

\begin{equation}
\label{eq19} \chi _{in} = \chi _{mb} + \lambda (\chi _{ms} - \chi
_{mb} ),\quad\quad (0 \le \lambda \le 1),
\end{equation}

\noindent where $\chi _{in} \equiv \sqrt {{r_{in} }
\mathord{\left/ {\vphantom {{r_{in} } M}} \right.
\kern-\nulldelimiterspace} M} $, $\chi _{ms} \equiv \sqrt {{r_{ms}
} \mathord{\left/ {\vphantom {{r_{ms} } M}} \right.
\kern-\nulldelimiterspace} M} $ and $\chi _{mb} \equiv \sqrt
{{r_{mb} } \mathord{\left/ {\vphantom {{r_{mb} } M}} \right.
\kern-\nulldelimiterspace} M} $ are all dimensionless radial
parameters. Inspecting equations (\ref{eq6}), (\ref{eq7}) and
(\ref{eq9}), we find that both the half-opening angle $\theta _M $
and the specific energy $E_{ms} $ are closely related to the
position of the inner edge of the disc, which is characterized by
$\chi _{ms} $. Therefore we expect that both the jet powers and
radio loudness of AGNs are affected by the parameter $\lambda $,
in terms of which the inner edge of the disc is determined.

Replacing the parameter $\chi _{ms} $ by $\chi _{in} $ in equation
(\ref{eq15}), we have the curves of jet power $\tilde {P}_{jet} $
versus $a_\ast $ for the given $\lambda $ and those of $\tilde
{P}_{jet} $ versus $\lambda $ for the given $a_\ast $ in the case
of the thick disc as shown in Figs.8a and 8b, respectively. The
power-law index $n = 6$ is assumed in the following calculations.

Similarly, replacing the parameter $\chi _{ms} $ by $\chi _{in} $
in equation (\ref{eq11}), we have the contours of radio-loudness
of AGNs in the parameter space consisting of $a_ * $ and $\lambda
$ as shown in Fig.9, where the value of $P / L$ in the case of the
thick disc is labeled beside each contour, and the thick solid
line labeled 0.001 is regarded as the dividing line between RQ and
RL AGNs (M97).

From equation (19) we find that the higher the value of $\lambda
$, the closer the inner edge of the disc to the marginally stable
orbit. Inspecting Figs.8 and 9, we find that both jet powers and
radio loudness of AGNs are indeed affected by the parameter
$\lambda $. The higher the value of $\lambda $, the greater the
jet powers and radio loudness of AGNs. These results imply that
these two quantities are rather sensitive to the position of the
inner edge of an accretion disc. Therefore we conclude that the
``spin paradigm'' for radio loudness of AGNs might be modified by
a scenario containing more than one physical parameter besides the
BH spin.

\noindent\textbf{Acknowledgments. }This work is supported by the
National Natural Science Foundation of China under Grant Numbers
10173004, 10373006 and 10121503.


\begin{thebibliography}{99}

\bibitem{1}{Abramowicz M. A., Jaroszynski M., Sikora M., 1978, A{\&}A 63, 209}

\bibitem{2}{Abramowicz M.A., Lasota J.P., 1980, Acta. Astro. 30, 35}

\bibitem{3}{Bardeen J. M., 1970, Nature, 226, 64}

\bibitem{4}{Beskin V. S., 1997, Physics Uspekhi, 40, 659}

\bibitem{5}{Blandford R. D., 1976, MNRAS, 176, 465}

\bibitem{6}{Blandford R. D., Znajek R. L., 1977, MNRAS, 179, 433}

\bibitem{7}{Blandford R. D., 1990, in Active Galactic Nuclei, ed. T. J.-L.
Courvoisier {\&} M. Mayor (Saas-Fee Advanced Course 20)
(Berling-Springer), 161}

\bibitem{8}{Cao X. W., Rawlings S., 2004, MNRAS, 349, 1419}

\bibitem{9}{Celotti A., Padovani P., Ghisellini G., 1997, MNRAS, 286,
415}

\bibitem{10}{Cirasuolo M., Celotti A., Magliocchetti, M., Danese, L., 2003, MNRAS, 346,
447}

\bibitem{11}{Kellerman K.I., Sramek R., Schmidt M., Shaffer D.B., Green
R.,1989, AJ, 98, 1195}

\bibitem{12}{Li L. X., 2002, A{\&}A, 392, 469}

\bibitem{13}{Macdonald D., Thorne K. S., 1982, MNRAS, 198, 345}

\bibitem{14}{Marchesini D., Celotti A., Ferrarese L., 2004, MNRAS, 351,
733}

\bibitem{15}{Moderski R., Sikora M., Lasota J.P., 1997, in
``\textit{Relativistic Jets in AGNs}'' eds.M. Ostrowski, M.
Sikora, G. Madejski {\&} M. Belgelman, Krakow, p.110 (M97)}

\bibitem{16}{Novikov I. D., Thorne K. S., 1973, in \textit{Black Holes}, ed.
Dewitt C

(Gordon and Breach, New York) p.345}

\bibitem{17}{Rees M. J., 1984, ARA{\&}A, 22, 471}

\bibitem{18}{Wang D. X., 1999, A{\&}A, 347, 1069}

\bibitem{19}{Wang D. X., Xiao K., Lei W. H., 2002, MNRAS, 335, 655
(W02)}

\bibitem{20}{Wang D. X., Lei W. H., Ma R. Y., 2003a, MNRAS, 342, 851
(W03a)}

\bibitem{21}{Wang D. X., Ma R. Y., Lei W. H., Yao G. Z., 2003b, ApJ, 595, 109
(W03b)}

\bibitem{22}{Wang J. M., Ho L. C., Staubert R., 2003, A{\&}A, 409,
887}

\bibitem{23}{Willem W., Wang T. G., Norbert S., Roberto V., 1997, MNRAS, 288,
225}

\bibitem{24}{Wilms J et al., 2001, MNRAS, 328, L27}

\bibitem{25}{Wilson A. S., Colbert E. J. M., 1995, ApJ, 438, 62}

\bibitem{26}{Xu C., Livio M., Baum S., 1999, ApJ, 118, 1169}



\end{thebibliography}
\end{document}